\documentclass[journal]{IEEEtran}
\usepackage{amssymb}
\usepackage{latexsym}

\usepackage{url}
\usepackage{xcolor}
\definecolor{newcolor}{rgb}{.8,.349,.1}

\usepackage{hyperref}

\usepackage[switch,pagewise]{lineno} 
\usepackage{makecell}
\usepackage{bm}
\usepackage{hhline}
\usepackage{longtable}
\usepackage{floatrow}
\usepackage{amsmath,amssymb,amsfonts}
\usepackage{algorithmic}
\usepackage{graphicx}
\usepackage{textcomp}
\usepackage{xcolor}
\usepackage{float}
\usepackage{subfig}
\usepackage{multirow} 
\usepackage{xcolor}
\usepackage{bm}
\usepackage{comment}
\usepackage{url}
\usepackage{hyperref}
\usepackage{longtable}
\usepackage[linesnumbered,ruled]{algorithm2e}
\usepackage{algorithmic}
\usepackage{paralist}
\hypersetup{
	colorlinks=false,
	pdfborder={0 0 0},
}

\usepackage{caption}
\usepackage{amsmath}

\begin{document}



\title{Aggressive saliency-aware point cloud compression}


\author{Eleftheria~Psatha, Dimitrios~Laskos, Gerasimos~Arvanitis,
and~Konstantinos~Moustakas,~\IEEEmembership{Senior~Member,~IEEE}
\thanks{E. Psatha, D. Laskos, G. Arvanitis, and K. Moustakas are with the Department of Electrical and Computer Engineering, University of Patras, Greece (e-mail: 
elefpsatha@gmail.com, dim-laskos@hotmail.com, arvanitis@ece.upatras.gr, moustakas@ece.upatras.gr)} 
}

\markboth{Journal of \LaTeX\ Class Files,~Vol.~14, No.~8, August~2022}%
{Shell \MakeLowercase{\textit{et al.}}: Bare Demo of IEEEtran.cls for IEEE Journals}

\maketitle
    



\begin{abstract}
The increasing demand for accurate representations of 3D scenes, combined with immersive technologies has led point clouds to extensive popularity. However, quality point clouds require a large amount of data and therefore the need for compression methods is imperative. In this paper, we present a novel, geometry-based, end-to-end compression scheme, that combines information on the geometrical features of the point cloud and the user’s position, achieving remarkable results for aggressive compression schemes demanding very small bit rates. After separating visible and non-visible points, four saliency maps are calculated, utilizing the point cloud’s geometry and distance from the user, the visibility information, and the user’s focus point. A combination of these maps results in a final saliency map, indicating the overall significance of each point and therefore quantizing different regions with a different number of bits during the encoding process. The decoder reconstructs the point cloud making use of delta coordinates and solving a sparse linear system. Evaluation studies and comparisons with the geometry-based point cloud compression (G-PCC) algorithm by the Moving Picture Experts Group (MPEG), carried out for a variety of point clouds, demonstrate that the proposed method achieves significantly better results for small bit rates.
\end{abstract}




\section{Introduction}
In recent years, 3D point clouds (PCs) have become a very popular immersive multimedia representation of static and dynamic 3D objects and scenes. They are widely used in various fields such as 3D scanning and modeling \cite{9531524}, industrial applications \cite{9899707}, bio-medical imagery \cite{7792613}, surveillance systems \cite{9884073}, autonomous navigation \cite{10044746}, and Virtual/Augmented Reality (VR/AR) \cite{9974380}. These types of representations can enable users to freely navigate in a fully immersive 3D environment interacting with different 3D objects. Unfortunately, such dense representations require a large amount of data, which are not feasible for transmission on today’s networks \cite{9882132}.
PCs, in comparison with 3D polygon meshes, provide a simpler, denser and more compact form that does not require topology information, and therefore, they are suitable for storage/transmission, as well as in low-level processing such as registration \cite{9459476}, segmentation \cite{9956014}, object detection \cite{9999269}. However, realistic PCs should be dense and so, they require a huge amount of memory or bandwidth for transmission. As a result, efficient compression methods for 3D PCs are mandatory in order to achieve high performance and visual accuracy in such cases.  
Although many 3D PC compression methods have been proposed over the years, there is still a lot of effort to be made in order to achieve sufficient compression ratios, especially in applications which focus on user interactive and realistic viewing experiences and thus demand relatively high resolution of PCs. 
The main idea behind this research is to compress a simplified version of the original 3D PC, that consists of different levels of resolution, based on the viewpoint and the geometric characteristics of the PC. We propose a method that highlights the most visually significant parts of the PC and compresses the position of each point based on its ``extended saliency" that combines the viewer's relative position and geometric saliency. The main contributions and the innovative parts of our approach are summarized below: 
\begin{itemize}
	\item We propose a novel end-to-end geometry compression scheme that consists of a visibility-aware simplification module, a multi saliency estimation module, a compression module and a post-processing module for PC reconstruction.
	\item We use an extended saliency metric of visual importance based on PC's geometry features and user's position.
	\item The visible points are compressed with a bit rate proportional to their saliency, while non-visible points are not considered during encoding and are decoded solely based on their connectivity information. 
	\item Extensive evaluation studies to examine important aspects of our approach in order to highlight its applicability for practical scenarios.
\end{itemize}
Experimental evaluation, carried out using a collection of different datasets, shows that the proposed saliency-aware quantization approach achieves significantly high compression ratios, preserving at the same time geometrically meaningful perceptible areas. These special characteristics make the method ideal for using it in applications that require extremely high compression rates and good perceptual accuracy at the same time.

The rest of this paper is organized as follows: Section 2 presents previous and related works in the area of PC compression. Section 3 describes in detail the proposed method. Section 4 presents the experimental results of our approach and we compare them with other State-of-the-Art (SoA) methods of the literature. Section 5 draws the conclusions, limitations of the method and future work.
\begin{figure*}
	\begin{center}
		\includegraphics[width=0.82\linewidth]{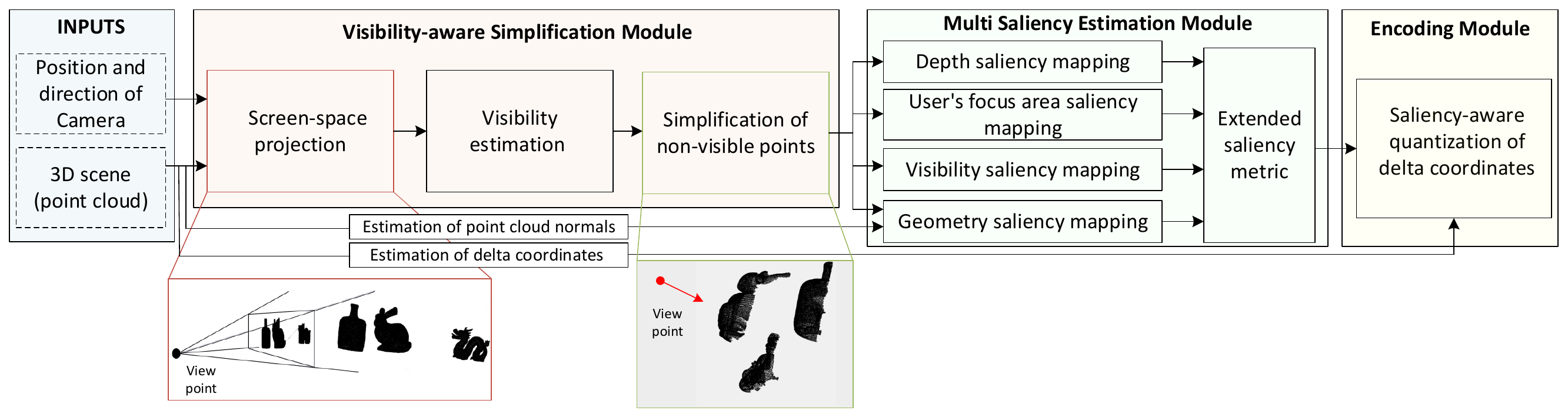}
	\end{center}
	\caption{Compression pipeline of the proposed methodology. The Multi-Saliency estimation module assigns a value to each visible point of the 3D scene indicating its perceptual significance. The encoding module compresses each point with a bit rate proportional to these values.}
	\label{pipeline}
\end{figure*}
\section{Previous work}
A PC is a set of points in three-dimensional space. Apart from its positional data, each point can also be associated with extra information including colors, normals, etc. Compression of 3D PCs has received significant attention during the past few years. Many approaches have been proposed in the literature to support efficient compression of PCs, including geometry coding \cite{8931233} (such as octree \cite{7924670,6224647}, kd-tree \cite{4061552}, spinning tree \cite{10.1145/1187112.1187277}, quadtree and binary-tree  approaches \cite{9105722} etc.), attribute coding \cite{8676054} (such as Graph Fourier Transform (GFT) \cite{Arvanitis_2020}, Karhunen-Loève transform (KLT) \cite{7914676}, Region-Adaptive Haar Transform (RAHT) \cite{7482691,8451367},
structured dictionary learning \cite{9204812}, etc.), or a combination of both \cite{9165178}.
The current SoA belongs to the Moving Pictures Expert Group (MPEG). There are 2 MPEG-PCC standards that are used in different scenarios, video-based PCC (V-PCC) 
for dynamic PCs and geometry-based PCC (G-PCC) for static PCs. The reference software for the aforementioned standards is TMC2 and TMC13 respectively \cite{mpeg}. Both V-PCC and G-PCC were based on traditional models such as octree decomposition, triangulated surface model, region-adaptive hierarchical transform, and 3D-to-2D projection \cite{wang2019learned}. 
V-PCC projects 3D points onto 2D planes and then uses existing video codec such as H.265/High-Efficiency Video Coding standard (HEVC) to compress the 2D planes. TMC13 is an octree-based geometry compression codec that can use trisoup surface approximations. It is equivalent to the combination of L-PCC (LIDAR PC compression for dynamic PCs) codec and S-PCC (Surface PC compression for static PCs) codec, previously used by MPEG. 
The compression is applied directly in the original 3D space, leading to acceptable quality results in both lossy and lossless intra-frame modes. Geometry and attribute information are encoded separately. 
There are 2 codecs available for the geometry analysis, an octree decomposition scheme and a trisoup (“triangle soup”) surface approximation scheme.
For attribute encoding, there are 3 codecs available: RAHT, Predicting Transform, and Lifting Transform. 
More details about V-PCC and G-PCC can be found at \cite{dumic2020point,8571288,graziosi2020overview}.   
Zhu et al. \cite{9057563} proposed a view-dependent DPC compression method specialized for networked applications that belong to the category of 3D-to-2D dimensional reduction and HEVC-based 2D video coding.
Gu et al. \cite{8816692} proposed a compression scheme for the attributes of voxelized 3D PCs. This method takes into consideration some special characteristics of the 3D PC, by voxelizing the 3D PC into equal size blocks and looking for irregular structures. Sun et al. \cite{8648155} 
proposed a lossless compression scheme based on PC clustering, taking advantage of a prediction technique, which takes into consideration the correlation of the distance information of points in order to remove spatial redundancies. Li et al. \cite{8803233} take into account only the rate instead of the rate-distortion cost, trying to overcome the problematic situations where some unoccupied pixels between different patches are compressed using almost the same quality as the occupied pixels, leading to waste of lots of bits since the unoccupied pixels are useless for the reconstructed PC.
Dricot and Ascenso \cite{8712753} proposed a hybrid PC compression approach that combines octrees with plane surfaces. More specifically, the octree partitioning is adaptive and also includes plane coding mode for leaf nodes at different layers of the octree. Tang et al. \cite{7924670} presented an approach of the octree coding algorithm that improves the stop condition so that the segmentation process stops dividing at the right depth, ensuring in this way an appropriate voxel size. Liu et al. \cite{9106052} proposed a coarse-to-fine rate control algorithm for region-based 3D PC compression. First, allocating the target bitrate between the geometry and color information, and then, optimizing, in turn, the geometry and color quantization steps. Mekuria et al. \cite{7786244,7434610} developed an octree-based intra- and inter-coding system. While Garcia and de Queiroz \cite{8451802} extended these approaches by designing a lossless intra-frame compression method applied to the PC geometry, where each octant in the octree is entropy-coded according to its father octant.
Shao et al. \cite{8305131} introduced a binary tree-based PC partition in order to achieve better energy compaction and compression efficiency, by using graph signal processing tools, like the graph transform with optimized Laplacian sparsity.
PC visibility has gained popularity over the last two decades. Many works distinguish visible and non-visible points from a viewpoint by reconstructing the surface, although they often require dense PCs and information about the normals. Katz et al. \cite{katz2007direct} estimated the visibility directly from point sets by spherical inverting every point and by calculating the convex hull of the inverted point set. Based on this approach, only the points that are lying on the convex hull are visible. Mehra et al. \cite{mehra2010visibility} enhanced this method to function well with noisy PCs as well. 
Nevertheless, despite the very good reconstruction results that the existing compression methods provide, they focus more on the coding of the whole 3D object and less on areas clearly perceptible by the observer so as to optimize compression and observable reconstruction quality. Additionally, for aggressive bit rates, the quality of the reconstructed PCs in these methods is relatively low.

\section{Saliency-aware Compression} 
Initially, we introduce the basic definitions and preliminaries related to 3D static PC processing. Then, we explain the proposed saliency-aware compression scheme in detail. A brief representation of the pipeline is illustrated in Figure \ref{pipeline}. In a nutshell, the process starts by estimating the non-visible vertices of the PC based on the user's position and viewpoint. Then, the hidden vertices are simplified and four separate saliency maps are estimated, depending on: 	1) Geometrical features, 2) Visibility, 3) Proximity between the user and the PC, 4) User's focus point.
Based on a combination of these maps, each point of the simplified PC is associated with a value that indicates its significance. Afterward, the point-cloud delta coordinates, which are calculated using an approximate connectivity graph, are scaled, quantized and eventually entropy coded. Each point is encoded with a different number of bits depending on its extended saliency. Finally, on the decoder side, we reconstruct the PC using the decoded delta coordinates.
\subsection{Basic Definitions}
In this work, we focus on PCs $\mathbf{P}$ consisting of $n$ vertices $\mathbf{v}$. A $\mathbf{P}$ can represent a 3D object or a scanned scene, consisting of different visible 3D objects. The $i$-th vertex $\mathbf{v}_i$ is represented by the Cartesian coordinates, denoted $\mathbf{v}_i = \left[x_i,\ y_i,\ z_i\right]^T, \ \forall \ i = 1,\cdots,n$. Thus, all the vertices can be represented by the matrix $\mathbf{V} = \left[\mathbf{v}_1,\ \mathbf{v}_2,\ \cdots,\mathbf{v}_n\right]\in \mathbb{R}^{3\times n}$. Each vertex $ \mathbf{v}_i\in \ \mathbf{V} $ is also represented by an outward unit normal $\mathbf{n}_i \ i = 1, \cdots,n$. Since connectivity information is not available, the computation of normals is achieved by local surface estimators.
The $k$ nearest neighbors of point $i$ are denoted by $\Psi_i^k$. Throughout the paper each neighboring point $j$ can be indicated through its vertex coordinates ($\mathbf{v}_j\in \Psi_i$) or, for simplicity, only through its index ($j \in \Psi_i$). The position of the user (or camera) in the 3D space of the input PC is symbolized by the point $ \footnotesize \mathbf{e} = \begin{bmatrix}
e_x & e_y & e_z
\end{bmatrix}^T,$ while the viewing direction by the vector $ \footnotesize \mathbf{r}= \begin{bmatrix}
r_x & r_y & r_z
\end{bmatrix}^T$.

\subsection{Offline Processing}
These paragraphs refer to offline processes that run only once for each PC, since they are independent of the location of the users and the direction of view.
\subsubsection{Estimation of $\delta$ Coordinates}
The Laplacian matrix $\mathbf{L}  \in \mathbb{R}^{n \times n}$ can be defined as: $\mathbf{L} = \mathbf{D} -  \mathbf{C} \label{eq:Ldef}$
where $\mathbf{C} \in \mathbb{R}^{n\times n}$ is an approximate adjacency (connectivity) matrix of the PC, with elements:
\begin{equation}
	\small
	\mathbf{C}_{\left(i,j\right)} = \left\{\begin{array}{ll}1& \text{if} \left(i,j\right) \in \Psi_i^k\\
		0 & \text{otherwise}   \end{array}\right.
	\label{adjacency}
\end{equation}
and $\mathbf{D}$ is a diagonal matrix with $\mathbf{D}_{(i,i)} = \left|\Psi_i^k\right|$. 
To mention here that we use the k-nearest neighbors (k-NN) algorithm to approximately estimate the connectivity between the neighboring points since we do not have explicit knowledge of the underlying manifold. So, the local neighborhoods around each point act like the PC's estimated faces. For that reason, the selected number of neighboring points $k_n$ is small, typically ranging from 5 to 10.
The differential or $\boldsymbol{\delta}$ coordinates of a PC are calculated as the difference between the coordinates of each vertex $\mathbf{v}_i$ and the barycenter of its $k$ nearest neighbours, according to the following equation \cite{9733486}: 
\begin{equation}
	\small
	\boldsymbol{\delta}_i = [\delta_x, \ \delta_y, \ \delta_z]^T = \mathbf{v}_i - \frac{1}{|\Psi_i^k|} \sum_{j \in \Psi_i^k}  \mathbf{v}_j, \ \	\Rightarrow
	\ \ \boldsymbol{\delta} =  \mathbf{L}\mathbf{V}
	\label{delta_alliws}
\end{equation}
\subsubsection{Estimation of Point Cloud Normals} 
Typically, the normal at a certain point is calculated as the perpendicular vector to the surface at that point. However, since the input vertices represent a set of point samples on the actual surface, we estimate the surface normals directly from the PC. More precisely, we use a rather classic method \cite{hoppe1992surface}, which specifies the neighbors of each point within a certain scale, and then applies PCA regression to estimate a tangent plane. For more robust estimations, a larger scale is usually preferred. In our experiments, we use a neighborhood of $k_n$ points for the local plane fitting.
\subsection{Visibility-aware Simplification}
This paragraph briefly presents the visibility estimation method and we propose a visibility aware simplification process of the PC.
\subsubsection{Projection to Screen Space}
Given the location $\mathbf{e}$ of the user and the view direction $\mathbf{r}$, each $i$ vertex $\mathbf{v}_i$ of the PC is projected into a two-dimensional pixel $ \footnotesize u_i = \begin{bmatrix}
x_i & y_i
\end{bmatrix}^T$. To that end, we apply a series of transformations to the PC's vertices, known as the geometric pipeline. These transformations are presented in detail in \cite{eberly20063d}. All parameters, such as the near and far planes of the view frustum, have been chosen to be compatible with the PC's dimensions in order to ensure that the entire PC is within the viewing frustum.
\subsubsection{Calculation of Visibility and Point Cloud Simplification}
The used visibility estimation algorithm \cite{biasutti} does not require dense PCs, uniform sampling or information about the normals and it achieves accurate results without reconstructing the surface.
Having the screen-space projection, we calculate the depth of each point. The depth of point $i$, notated as $d_i$, is the Euclidean distance between point $i$ and the user's position $\mathbf{e}$. Using the k-NN algorithm, sets of $k_a$ nearest points with Euclidean distance in screen-space are computed. An operator $a \in \ [ 0, 1 ]$ is calculated for each point according to:
\begin{equation}
	\small
	a_i =\exp\biggl(-\dfrac{(d_i-{d_i}^{min})^2} {({d_i}^{max}-{d_i}^{min})^2}\biggr), \  \ \forall \ i = 1, \cdots, n
	\label{a_operator}
\end{equation}
where ${d_i}^{min}$ and ${d_i}^{max}$ are the minimum and maximum depths in $i$’s neighborhood respectively. Finally, the visibility of each point is determined by setting a threshold to the value of $a_i$. The threshold that we used is $a_{threshold}=a_{mean}$.
Points having $a_i$ below the value $a_{threshold}$, are regarded as non-visible from the point of view of the observer, and thus their simplification does not make a significant perceptually difference for the user. So, only the visible part of the PC will be assigned quantization bits in the next steps.

\subsection{Multi-saliency Estimation}


\subsubsection{Geometric Saliency}
\label{spectral_analysis}
Saliency detection, which aims to detect geometric features in 3D scenes, has been a hot topic in recent years and several feature descriptors have been created \cite{Han20183DPC}. Motivated by the fact that geometric features, like high curvature regions, corners, and edges, usually convey important visual information, we estimate a saliency map based on the geometric importance of each point. More specifically, we assume that areas with high-frequency spatial information are more perceptually significant and they must be preserved in contrast to flat areas. In our proposed geometric saliency scheme, we combine an eigenvalue-based step which extracts saliency features by decomposing local covariance matrices defined in small regions around each point of $\mathbf{P}_v$ \cite{9120202}, and a step that uses the so-called Darboux frame as a local curvature metric for each region. Consider that for each point $i$ of the $\mathbf{P}_v$ we can create a matrix $\mathbf{E}_i \ \in \ \mathbb{R}^{3 \times(k_g+1)}$ that consists of the normals of its corresponding $k_g$ nearest neighbors that were estimated during the offline processing. This matrix is formulated according to:
\begin{equation}
	\small
	\mathbf{E}_{i} = \begin{bmatrix}
	{n}_{ix} \ {n}_{ix_1} \ {n}_{ix_2}  \dots   {n}_{ix_{k_g}} \\
	{n}_{iy} \ {n}_{iy_1} \ {n}_{iy_2}  \dots   {n}_{iy_{k_g}} \\
	{n}_{iz} \ {n}_{iz_1} \ {n}_{iz_2}  \dots   {n}_{iz_{k_g}} \\
	\end{bmatrix}, \  \ \forall \ i = 1,\cdots, {n}_v
	\label{NN1}
\end{equation}
Then, the matrix $\mathbf{E}_{i}$ is used for the estimation of the local covariance matrices $\mathbf{R}_i  = \mathbf{E}_{i} \mathbf{E}_{i}^{T} \in \mathbb{R}^{3\times 3}$  
and next, the matrix $\mathbf{R}_{i} = \mathbf{U} \mathbf{\Lambda} \mathbf{U}^{T}$ is decomposed to the matrix $\mathbf{U}$ of the eigenvectors, and the diagonal matrix $\mathbf{\Lambda} = \text{diag}(\lambda_{i1}, \lambda_{i2}, \lambda_{i3})$ of the eigenvalues $\lambda_{ij}, \ \forall \ j=\{1,2,3\}$. The saliency $s_{11i}$ based on the geometry of each vertex is denoted as the value given by the inverse $l^2$-norm of the corresponding eigenvalues \cite{8283576}:
\begin{equation} 
	\small
	s_{11i} =  \frac{1}{\sqrt{\lambda_{i1}^2 + \lambda_{i2}^2 + \lambda_{i3}^2}}, \ \ \forall \ i = 1,\cdots, {n}_v 
	\label{spectral_saliency}
\end{equation}
From Eq. \eqref{spectral_saliency}, we can observe that large values of the term $\footnotesize \sqrt{\lambda_{i1}^2 + \lambda_{i2}^2 + \lambda_{i3}^2}$ correspond to small saliency features indicating that the point lies in a flat area, while small values correspond to large saliency values, characterizing the specific point as a feature. 
In order to further highlight geometrical significance in neighborhoods around the the most salient points previously detected, we extract  features of high curvature values from such neighbourhoods. To be more specific, we exploit the most salient vertices of Eq. \eqref{geom_saliency_final}, which are denoted by $\mathbf{v}_i \in \mathbf{P}_s:\overline{s}_{11i}>s_0$, where $\mathbf{P}_s \subseteq \mathbf{P}_v$. The threshold value of $s_0$ is set such that only the largest saliency values are preserved. To efficiently obtain more informative features, we propose the computation of the Darboux frame for local regions defined by each point $\mathbf{v}_i$ of $\mathbf{P}_s$ and its $k_g$ closest neighbors. The Darboux frame is a canonical moving frame that consists of three orthonormal vectors $\mathbf{g}_1$, $\mathbf{g}_2$, $\mathbf{g}_3$ based at a point $\mathbf{v}$ and it is considered to be a local representation of the surface. Consider that for every pair of points $\mathbf{v}_i \in \mathbf{P}_s$ and $\Psi_i^k$ in the k-neighbourhood of $\mathbf{v}_i$, we select a source $\mathbf{v}_{si}$ and a target $\mathbf{v}_{ti}$ point. The source is the one having the smaller angle between the associated normal vector and the line connecting the points \cite{rusu2008persistent}. The vectors of the Darboux frame constructed at point $\mathbf{v}_{si}$ are computed as:
\begin{equation}
	\small
		\mathbf{g}_{1i} = \mathbf{n}_{si}, \ \ 
		\mathbf{g}_{2i} = \mathbf{g}_{1i} \times \frac{ (\mathbf{v}_{ti} - \mathbf{v}_{si} )} {{\lVert \mathbf{v}_{ti} - \mathbf{v}_{si}  \rVert}_2}, \ \ 
		\mathbf{g}_{3i} =\mathbf{g}_{1i} \times \mathbf{g}_{2i}
	\label{di_frame}
\end{equation}
The aforementioned vectors give us additional information about every local surface region, so we create for each one of them a matrix $\mathbf{G}_{ji}\ \in \ \mathbb{R}^{3 \times(k+1)}$ that consists of the Darboux frame's vectors between the point i and its corresponding $k_g$ nearest neighbours, according to:
\begin{equation}
	\small
	\mathbf{G}_{ji} = \begin{bmatrix}
		g_{{j}_{ix}} & g_{{j}_{ix_1}} & g_{{j}_{ix_2}} & \dots  & g_{{j}_{ix_{k_g}}} \\
		g_{{j}_{iy}} & g_{{j}_{iy_1}} & g_{{j}_{iy_2}} & \dots  & g_{{j}_{iy_{k_g}}} \\
		g_{{j}_{iz}} & g_{{j}_{iz_1}} & g_{{j}_{iz_2}} & \dots  & g_{{j}_{iz_{k_g}}} 
	\end{bmatrix} 
	\label{NN2}
\end{equation}
$\forall \ \ j  =1, 2 ,3$ and $  i = 1,\cdots, {n}_s$. We continue by estimating in a similar way the local covariance matrices and corresponding eigenvalues for each orthonormal vector of the Darboux frames. For each point $\mathbf{v}_i \in \mathbf{P}_s$, we define a local curvature metric $c_i$ around its region using the eigenvalues of the vectors  $\mathbf{g}_1$, $\mathbf{g}_2$, $\mathbf{g}_3$:
\begin{equation}
	\small
	c_{i}  = {\sum_{j=1}^{3} \frac{1}{\sqrt{\lambda_{\mathbf{g}_{ji_1}}^2 + \lambda_{\mathbf{g}_{ji_2}}^2 + \lambda_{\mathbf{g}_{ji_3}}^2}}}, \ \ \
	\forall \ \ i = 1,\cdots, {n}_s
	\label{NN3}
\end{equation}
Flat regions correspond to small values of $c_i$, while high curvature regions correspond to larger values, it has been observed. The high-curvature points are considered geometrically significant and must therefore be preserved. The saliency $s_{12i}$ based on the aforementioned metric is defined below:
\begin{equation}
	\small
	s_{12i}  = \text{max}({c_i}) - \frac{1}{1 - e^{c_i}}, \ \ 
	\forall \ i = 1,\cdots, {n}_s
	\label{NN3}
\end{equation}
$s_{12i}$ amplifies the importance of geometric features and also leads to more salient features being detected around each neighbourhood of interest that was extracted by the first step (meaning that they transition into a more salient category). The above equation presents the final saliency map that detects features based on local curvature and geometric saliency:
\begin{equation}
	\small
	s_{1i} = \left\{\begin{matrix}
		s_{12i} , & \text{if} & s_{11i}>s_0
		\\ s_{11i} , & \text{otherwise}  
	\end{matrix}\right. , 
	\ \ \forall \ \   i = 1,\cdots, {n}_v
	\label{EQ}
\end{equation}
The final geometry saliency is normalized in the range $[0,1]$:
\begin{equation}
	\small
	\overline{s}_{1i} = \frac{s_{1i} - \text{min}(s_{1i})}{ \text{max}(s_{1i}) - \text{min}(s_{1i})}, \ \ \forall \ i = 1,\cdots, {n}_v
	\label{geom_saliency_final}
\end{equation}
\begin{figure}[!htbp]
	\begin{center}
		\includegraphics[width=0.65\linewidth]{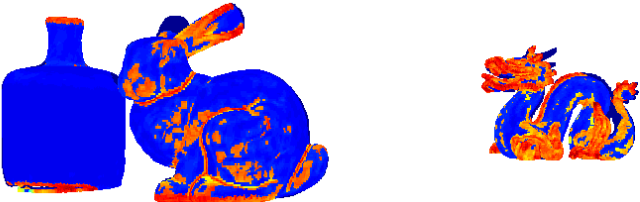}
	\end{center}
	\caption{Heatmap visualization of the saliency map based on vertices geometry.}
	\label{geometry}
\end{figure}
Fig. \ref{geometry} illustrates an example of a heatmap that visualizes the saliency map of the vertices based on their geometry and curvature. As we can see, high-frequency spatial areas, like sharp corners, are represented by deep red color while vertices lying in local flat areas represented by deep blue color.
\subsubsection{Visibility Saliency}
\begin{figure}[!htbp]
	\begin{center}
		\includegraphics[width=0.6\linewidth]{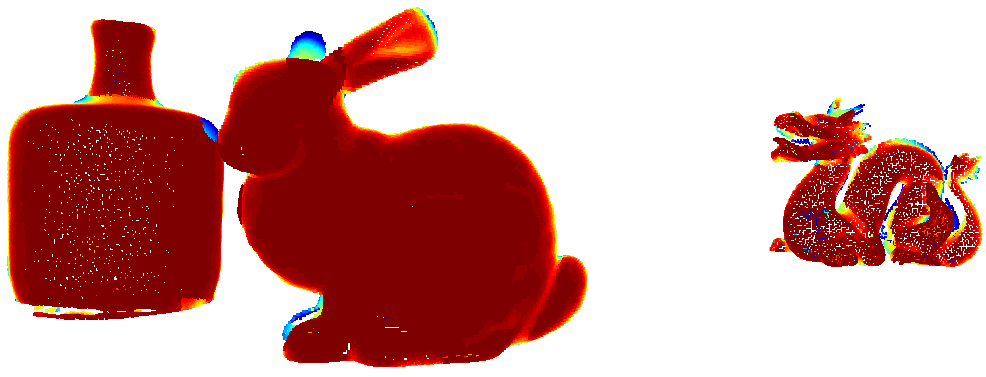}
	\end{center}
	\caption{Heatmap visualization of the saliency map based on vertices visibility.}
	\label{visibility}
\end{figure}

The visibility operator $a$, presented in Eq. \eqref{a_operator}, is a  metric defining the confidence that a point i is not occluded. More specifically, the closer to 1 is $a_i$ the more certain we are that the point i is visible. We create a saliency map based on this operator, assuming that points with high $a$ values are visually more salient.
So, the saliency $s_2i$ of the point $i \in\ P_v$ that is based on the visibility operator will be denoted as:
\begin{equation}
	\small
	s_{2i} =  \ a_i, \ \ \forall \ i = 1,\cdots, {n}_v 
	\label{visibility_saliency}
\end{equation}
The geometry of the less salient points of the visible part (where $a$ is close to $a_{threshold}$) will be encoded with fewer bits.  Fig. \ref{visibility} presents a heatmap visualization of the visibility based saliency map. The visible points are represented by dark red colour, while the fully occluded points by dark blue color.

\subsubsection{Depth from Viewpoint Saliency}
\begin{figure}[!htbp]
	\begin{center}
		\includegraphics[width=0.65\linewidth]{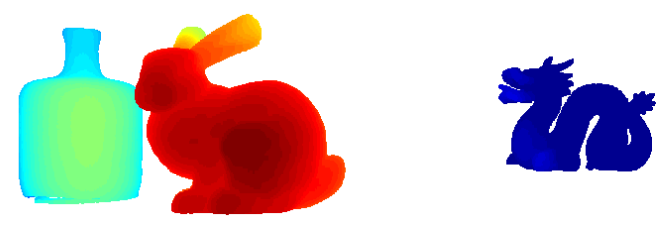}
	\end{center}
	\caption{Heatmap visualization of the depth saliency map.}
	\label{depth_saliency}
\end{figure}

Due to the restricted depth of field of the human eye, objects that are far from the viewpoint tend to appear blurred. In detail, visual acuity is reduced at the parts of the field of view with higher depth values. Motivated by this fact, we calculate a saliency map based on depth difference from the point of view. The user will not notice the visual difference in quality reduction.
The depth $d_i$ of each $i$ point has already been calculated during the simplification step. We propose using a modified version of a depth map that was initiated by \cite{lee2008real}. The saliency $s_{3i}$ of the point $\mathbf{v}_i \in\ \mathbf{P}_v$ is derived from normalising and transforming the depth values as:
\begin{equation}
	\small
	s_{3i} = 1 - \frac{d_i - z_{near}}{z_{far} - z_{near}},   \ \forall \ i = 1,\cdots, {n}_v  
	\label{depth_saliency_final}
\end{equation}
where $z_{near}$ and $z_{far}$ are the distances of the near and far clip planes of the viewing frustum. This transformation is applied so that the most salient points would be those that are closest to the point of view. In  Fig.  \ref{depth_saliency}, we present a  heatmap visualization of the depth-based saliency map. The points closer to the user's position are presented in dark red color, while the points far from the user are presented in dark blue color.
\subsubsection{Focus Point Saliency}
\begin{figure}[!htbp]
	\begin{center}
		\includegraphics[width=0.65\linewidth]{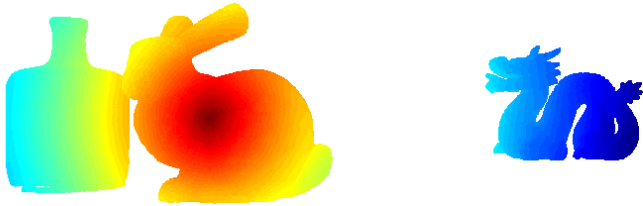}
	\end{center}
	\caption{Heatmap visualization of the saliency map based on the users' focus.}
	\label{focus_saliency}
\end{figure}

The point that the users' eyes are looking at is called as ``focus" point and can be derived from their position and viewing direction. It is known that humans perceive more details close to the focus point than at its periphery. In this section, we estimate a saliency map by incorporating a peripheral blur effect that progressively reduces the quality of the scene at points located at a certain distance focus point. This effect is based on \cite{hillaire2008depth} and it is independent of the depth-of-field. The quality reduction in the periphery adds a realistic sensation and is not noticeable by the user. The saliency $s_{4i}$ of the point $\mathbf{v}_i \in\ \mathbf{P}_v$, is defined as:
\begin{equation}
	\small
	s_{4i} = 1 - \sqrt{\frac{1}{\mathbf{r}\cdot p_i}}^m,  \forall \ i = 1,\cdots, {n}_v  
	\label{focus_s}
\end{equation}
where $\mathbf{r}$ is the user's viewing direction and $p_i$ is the normalised vector from point $\mathbf{v}_i$ towards the user's position. The power $m$ defines the size of the area around the focus point where the quality is almost unaffected. For our experiments we use $m=1$.
Finally, $s_{4i}$ is normalized in the range $[0,1]$ according to:
\begin{equation}
	\small
	\overline{s}_{4i} =   \frac{s_{4i} - \text{min}(s_{4i})}{ \text{max}(s_{4i}) - \text{min}(s_{4i})}, \ \ \forall \ i = 1,\cdots, {n}_v
	\label{focus_saliency_final}
\end{equation}
The most salient points that will be preserved with more geometrical accuracy are the ones with the smallest distance from the users' focus point. In  Fig.  \ref{focus_saliency}, we present a heatmap  visualization of the focus point based saliency map. The focus point is located at the center of the red circles.
\subsubsection{Extended Saliency Metric}
\begin{figure}[!htbp]
	\begin{center}
		\includegraphics[width=0.65\linewidth]{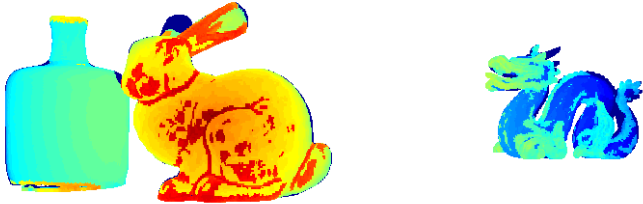}
	\end{center}
	\caption{Heatmap visualization of the final saliency map.}
	\label{total}
\end{figure}

The aforementioned saliency maps, based on a different strategy, indicate which points will be encoded with more or less geometric precision. An  extended  saliency  metric that combines the above maps is calculated as the sum of 2 functions:
\begin{equation}
	\small
	s_{i} =   f ( \overline{{s}_{1i}} , s_{2i} ) + g ( s_{3i} ,\overline{{s}_{4i}} ),  \ \ \forall \ i = 1,\cdots, {n}_v  
	\label{total_saliency_general}
\end{equation}
where $f$ is a function that calculates the visibility and geometry contribution to the final saliency map and the function $g$  calculates the depth and focus contribution. Visibility and geometry are objective measures of saliency and their contribution to the final map should therefore be greater compared to focus and depth. The last two are subjective, as the amount of visual acuity loss varies from user to user in both situations. Choosing carefully the aforementioned functions is necessary because too much emphasis on focus and depth maps could lead to perceptually significant quality reductions.
The final salience map is an indicator of the overall visual significance of each point in $\mathbf{P}_v$. The points with higher saliency values are preserved without any loss of precision as they are considered visually important, while the detail of the points with lower saliency is decreased without the user perceiving the visual difference.
In our experiments, we chose a linear combination of the individual saliency maps. To be more precise the extended saliency metric will be calculated as the weighted average of those maps:
\begin{equation}
	\small
	s_{i} =   \frac{ w_1 \overline{s}_{1i} +  w_2 s_{2i} +  w_3 {s}_{3i} +  w_4 \overline{s}_{1i} }{w_1 + w_2 + w_3 + w_4}, \ \ \forall \ i = 1,\cdots, {n}_v  
	\label{total_saliency}
\end{equation}
Each saliency has been normalized within a range of $[0,1]$, except for the visibility-based saliency that was already within that range. The corresponding weights ${w_{j}, \ \  \forall \ j = 1,\cdots, 4}$ can be tuned to emphasize one approach or the other. Further experiments will be held in order to study the contribution of each saliency to the quality of the decompressed PC and to optimally estimate these weights. In Fig. \ref{total}, we present a heatmap visualization of the final saliency map. In this particular visualization the effect of each different method to is equal on the final saliency map. The visually important points are represented with deep red colour, while the visually unimportant points are represented by dark blue color.

\subsection{Saliency-aware Point Cloud Encoding}
In the proposed saliency-aware PC compression scheme, we are able to progressively encode geometry information allowing the generation of a sequence of levels of detail. The main idea is that we allocate a different number of bits for the encoding, according to the overall extended saliency of each point in $\mathbf{P}_v$. The non-visible points are encoded with zero bits which means that none of their geometrical accuracy is preserved. We choose to apply our compression scheme to the model's $\boldsymbol{\delta}$ coordinates because it is proven that their quantization yields a small visual error, in contrast to the standard Cartesian coordinate quantization \cite{sorkine2003high}.
Based on the previously calculated $\delta$ coordinates, we set to zero the coordinates of the non visible part and we generate the matrix $\boldsymbol{\delta}  = \left[\delta_{1},\ldots, \delta_{n}\right]$:
\begin{equation}
	\small
	\boldsymbol{\delta}_{i} = \left\{\begin{array}{ll}
		0, & \text{if} \ \mathbf{v}_i \ \text{is non visible} \\
		\delta_i,& \text{otherwise}
	\end{array}\right., \ \ \forall \ i = 1,\cdots, n
	\label{delta}
\end{equation}
For the quantization of the above matrix we use a rather simple but effective quantization function. To be more precise, we downscale the $\delta$-coordinates by multiplying each $\delta_i$ with a scaling factor that is based on the overall saliency value of each point  and then we round the result to the closest integer coordinates, according to:
\begin{equation}
	\small
	\boldsymbol{\tilde{\delta}}_{i} = \left\{\begin{array}{ll}
		0, & \text{if} \ \mathbf{v}_i \ \text{is non visible} \\
		\text{round}( s_{thresh} s_i   \delta_i), & \text{otherwise}
	\end{array}\right. 
	\label{quant-delta}
\end{equation}
, $\forall \ i = 1,\cdots, n$ and $s_{thresh} > 0$. The scaling threshold, which is multiplied with each $\boldsymbol{\delta}_i$, is a constant whose value determines the number of bits used to encode each point. Greater values of $s_{thresh}$ allow higher percentages of geometric accuracy to be preserved, while very small values lead to zero $\boldsymbol{\delta}_i$ coordinates after rounding. For a given $s_{thresh}$, by multiplying each point with its overall saliency value we allocate more bits for the encoding of the visually important points. 
In order to reduce the magnitude of the transformed quantization errors, we quantize a set of known anchor points along with the input PC. The anchor points are uniformly distributed on the model surface and denoted by $\footnotesize \mathbf{v}_c = \mathcal{Q}([\mathbf{v}_{i_1},\cdots,\mathbf{v}_{i_{k_c}} ])$ where $i_{k_c}$ is the vertex index and $k_c$ corresponds to the $1\%$ of the total number of vertices $n$. 
After quantization, the matrix $\footnotesize \widetilde{\boldsymbol{\delta}}  = \left[\widetilde{\boldsymbol{\delta}}_{1},\ldots, \widetilde{\boldsymbol{\delta}}_{n+ k_c}\right]$ is compressed using an entropy encoder. For our experiments, we used an arithmetic encoder. The compression ratio of the anchor encoding is insignificant since they correspond to a very small percentage of the PC. We also compress each additional matrix that is needed on the decoder side for the reconstruction of the PC. The saliency values $s = [ s_{thresh} s_1 ,\ldots,  s_{thresh} s_{n} ]$ $ \forall \ i = 1,\cdots, n $  are arithmetically encoded and the Laplacian matrix is encoded using an efficient connectivity compression method \cite{ying2010edgebreaker}.
\begin{algorithm}
	\SetKwInOut{Input}{Input}
	\SetKwInOut{Output}{Output}
	\Input{Unorganized 3D point cloud  $\mathbf{P} \ \in \ \mathbb{R}^{n \times 3}$;}
	\Output{Reconstructed 3D point cloud $\mathbf{\bar{P}} \ \in \ \mathbb{R}^{n \times 3}$;}{
		Find the $n_v$ visible vertices based on $a$ via Eq. \eqref{a_operator};\\
		\For{$i = 1, \cdots, n_v$}{
			Estimation of $s_{1i}$ (geometry-based) via Eqs. \eqref{NN1}-\eqref{geom_saliency_final}; \\
			Estimation of $s_{2i}$ (visibility-based), via Eq. \eqref{visibility_saliency}; \\
			Estimation of $s_{3i}$ (depth-based), via Eq. \eqref{depth_saliency_final};\\
			Estimation of $s_{4i}$ (user's focus-based), via Eq. \eqref{focus_s};\\
		}
	}
	Estimate an extended saliency metric via Eq. \eqref{total_saliency};\\
	Estimate the delta coordinates via Eq. \eqref{delta_alliws};\\
	Compression giving different bits per vertices based on their saliency via Eq. \eqref{quant-delta};\\
	Reconstruction by solving the linear system of Eq. \eqref{lsm};
	\caption{Saliency-aware compression of PCs} \label{al:pipeline_AOI}
\end{algorithm}
\subsection{Decoding and Reconstruction}
The decoder decompresses the geometry, connectivity and saliency data and  the $\delta$ coordinates are scaled back to their original size by: $\boldsymbol{\widetilde{\delta}}_{ri} = \frac{\widetilde{\boldsymbol{\delta}_i}}{s_i}, \ \
\forall \ i = 1,\cdots, n$. 
Finally, reconstruction of the 3D PC is performed by solving the following sparse linear system:
\begin{equation}
	\small
	\left[\begin{array}{cc}
		\mathbf{L}\\
		\mathbf{I}_{k_c}
	\end{array}\right]\mathbf{v} = \left[\begin{array}{cc}
		\boldsymbol{\boldsymbol{\widetilde{\delta}_r }}\\
		\mathbf{v}_c\end{array}\right]
	\label{lsm}
\end{equation}
where $\mathbf{I}_{k_c} \in \mathbb{R}^{{k_c}\times n}$ is a sparse matrix with ones at the $i_{k_c}$ indices where the vertices $\mathbf{v}_c$ lie and zeros anywhere else, so that $\mathbf{v}_c = \mathbf{I}_{k_c}\mathbf{v}$ (anchor points). 
Since the non-visible points were assigned zero bits during encoding, their reconstruction is solely based on the PC's Laplacian matrix.
Algorithm \ref{al:pipeline_AOI} summarizes the most important steps of our approach.

\section{Experimental Results and Analysis}
In this section, we will evaluate the quality of our proposed compression scheme for static PCs and compare its performance with the MPEG'S G-PCC, a SoA compression standardization as stated above. 


\subsection{Datasets}
\begin{figure}[!htbp]
	\centering
	\includegraphics[width=0.1\linewidth]{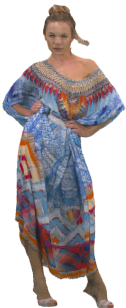}
	\includegraphics[width=0.1\linewidth]{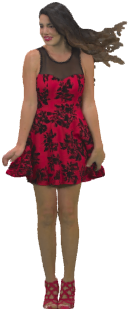}
	\includegraphics[width=0.08\linewidth]{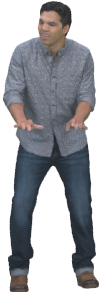}
	\includegraphics[width=0.11\linewidth]{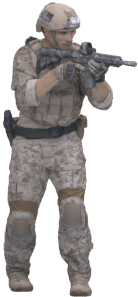}
	\includegraphics[width=0.12\linewidth]{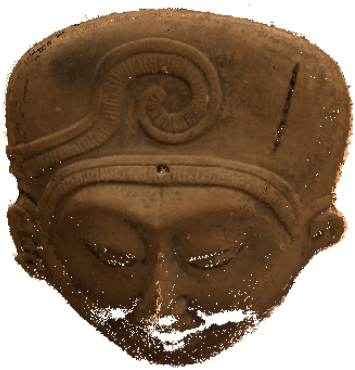}
	\includegraphics[width=0.063\linewidth]{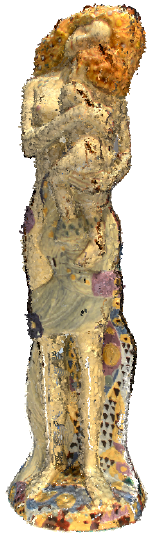}
	\includegraphics[width=0.16\linewidth]{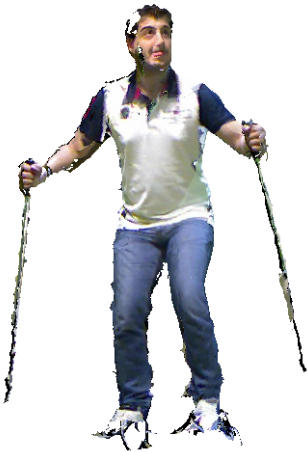}
	\includegraphics[width=0.08\linewidth]{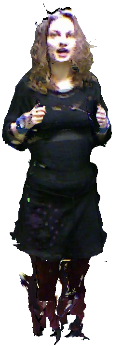}
	\includegraphics[width=0.4\linewidth]{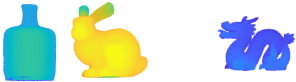}
	\caption{PCs used in experiments.} 
	\label{fig:pointClouds}
\end{figure}

We used multiple static PCs of different structures, complexities and properties in order to assess the quality of our proposed compression scheme. We used some frames from the dynamically acquired PCs from 8i Voxelized full Bodies dataset \cite{d20178i}, which have smooth and complete surfaces. These PCs were also chosen by MPEG for the current PCC standardization efforts according to \cite{perry2020jpeg}. We also tested  VCL/ITI's datasets (https://vcl.iti.gr/dataset/reconstruction) of multiple Kinect-based 3D reconstructed meshes based on \cite{alexiadis2012real}. We only used the 3D coordinates of this dataset and ignored the connectivity information that was also included. These models are affected by noise and consist of many holes and irregularities. We also chose 2 inanimate objects with a sparse but more precise voxel distribution from the MPEG database. For testing purposes, we have obtained these models by combining the non-overlapping patches of each object that have been generated and made available at \cite{guarda2020constant}. Finally, we constructed static scenes consisting of multiple models by using several PCs from \cite{8463406} dataset that is available publicly online (https://mmspg.epfl.ch/reconstructed-point-clouds-results).
The aforementioned PCs of our experimental test set are shown in Fig. \ref{fig:pointClouds} and detailed information about each one of them is presented in Table~\ref{table:datasets}. Most of these PCs were already in a voxelized form so we voxelized the remaining in order to be consistent with the required input form of G-PCC. This is necessary since G-PCC expects geometry to be expressed with integer precision, so all content must be quantized or voxelized before encoding.

\begin{table}[h]
	\centering
	\begin{tabular}{|c | c | c | c |} 
		\hline 
		\thead{ \textbf{Point} \textbf{Cloud}}	& \thead{ \textbf{Voxel}  \textbf{Depth}} & \thead{\textbf{Frame}} & \thead{\textbf{Vertices}} \\  
		\hline \hline
		\small Long dress & \small 10 & \small  1300 & \small 857,966 \\  
		\hline
	\small	Red and black & \small 10 & \small 1450 & \small 729,133 \\ 
		\hline
	\small	Loot & \small 10 & \small 1200 & \small 805,285 \\ 
		\hline
	\small	Soldier & \small 10 & \small 690 & \small 1,089,091\\ 
		\hline
	\small	Egyptian Mask & \small 12 & \small - & \small 274,432\\
		\hline
	\small	Statue Klimt & \small 12 & \small - & \small 499,712\\
		\hline
	\small	Skiing3-Zippering & \small 9 & \small 111 & \small 174,144\\ 
		\hline
	\small	S17\_5KW\_Xenia-Zippering & \small 9 & \small 1580 & \small 81,028\\ 
		\hline
	\small	Static scene & \small 10 & \small - & \small 94,698\\
		\hline
	\end{tabular}
	\caption{Datasets information.}	
	\label{table:datasets}
\end{table}

\subsection{Evaluation Metrics}
This section presents the SoA PSNR metrics used for the geometry quality evaluation defined by the MPEG \cite{schwarz2018common}. The original PC $V_{or} = \{(v_i): i = 0,..,K-1\}$ is a set of $K$ points without any relevant order. 
The decoded PC $V_{deg} = \{(v_i): i = 0,..,N-1$ consists of N points, typically $N<K$. 
The root mean square of the distances between the points of the two PCs (point-to-point metric \cite{8296925}) $d_{rms}$ is defined by the following equation where $v_{deg}^{nn}$ is the nearest point of $V_{deg}$ to the points of $V_{or}$.
\begin{equation}
	\small
	d_{rms}(V_{or},V_{deg}) = \sqrt {\dfrac{1}{K}\sum_{v_l\in{V_{or}}}[v_l - v_{deg}^{nn}]^2} 
\end{equation}
In a similar way the point-to-plane metric is defined by calculating the projection between each point's error vector $E$ along the normal vector $N$ of their underlying surface. 
\begin{equation}
	\small
	d_{p2plane}(V_{or},V_{deg}) = \sqrt {\dfrac{1}{K}\sum_{v_l\in{V_{or}}}(E(V_l,V_{or}) \cdot N_{or})^2} 
\end{equation}
For both of the metrics, we calculate their symmetric distances:
\begin{equation}
	\small
	d_{rms}^{sym}(V_{or},V_{deg}) = max(d_{rms}(V_{or},V_{deg}),d_{rms}(V_{deg},V_{or})) 
\end{equation} 
\begin{equation}
	\small{
		d_{p2plane}^{sym}(V_{or},V_{deg}) = max(d_{p2plane}(V_{or},V_{deg}),d_{p2plane}(V_{deg},V_{or})) 
	}
\end{equation} 
The geometry PSNR ratio can be computed both for the point-to-point and point-to-plane metrics \cite{8296925} using $d_{rms}^{sym}$ or $d_{p2plane}^{sym}$ respectively. 
\begin{equation}
	\small{
		bandwidth = max((x_{max}-x_{min}),(y_{max}-y_{min}),(z_{max}-z_{min}))
	}
\end{equation}
\begin{equation} 
	\small
	\label{eq:psnr}
	psnr_{geom} = 10\log_{10}\dfrac{\|bandwidth_{V_{deg}}\|^2_2}{(d_{rms/p2plane}^{sym}(V_{or},V_{deg}))^2 }
\end{equation} 
Metric (\ref{eq:psnr}) is referred to as D1 and D2 when calculated with point-to-point distance ($d_{rms}^{sym}$) and point-to-plane distance  ($d_{p2plane}^{sym}$) respectively.
Bjøntegaard model \cite{bjontegaard2001calculation,bjontegaard2008improvements} was also used to calculate the Bjøntegaard delta PSNR (BD-PSNR) which corresponds to the average PSNR difference for the same bit rate.

\subsection{Parameter Adjustment}
In this paragraph, we will present and justify the selection of parameter values that are fixed through the steps of the proposed methodology in order to provide reproducible results.
For the local surface fitting that is conducted for both the estimations of the PC's connectivity and normals, $k_n$ is a small number that depends on the density of the PC. For very small values of $k_n$, the estimated surface is affected by data noise and for large values, the k-neighbors become less localized and the resulting surface is not precise. To find the optimal number of $k_n$ for each model, we performed a grid search in the range $[3 - 15]$ using the aforementioned D1 and D2 metrics for the evaluation of the reconstructed PC. The optimal value is $k_n = 6$ for the entire dataset except for the Egyptian mask and the Klimt statue, which are sparser, and the best reconstruction is achieved by $k_n=5$.
Although users can be placed anywhere in the 3D scene of points ${v}_i = \left[x_i,\ y_i,\ z_i\right]^T \ \forall \ i = 1,\cdots,n$, we have chosen their position and viewing direction so that the whole scene is within the field of view. To be more specific, for each PC centered at (0,0,0) the user was located at $ \footnotesize e = \begin{bmatrix}
0 & 0 & 2 max{z}
\end{bmatrix}^T$
and the viewing direction was $\footnotesize r = \begin{bmatrix}
0 & 0 & - 2 max{z}
\end{bmatrix}^T$.
The $k_a$ neighboring points in screen space used for visibility estimation in Eq. \eqref{a_operator} and the $k_g$ neighboring vertices, lying in a patch area that is used for the estimation of matrices $\mathbf{E}$ and $\mathbf{G}$ in Eqs. \eqref{NN1}, \eqref{NN2} during geometry saliency estimation, are 2 parameters that play an important role in the quality of the final reconstructed model. However, the optimal value for each parameter can vary from model to model as they have different properties, structure, and density. To avoid exhaustive searching for optimal values for each model separately, we suggest using fixed values that provide good reconstruction results in most cases. For our experiments we have set $k_a = 125$ and $k_g  =  25$.
\begin{figure}[!htbp]
	\centering
	\includegraphics[width=0.95\linewidth]{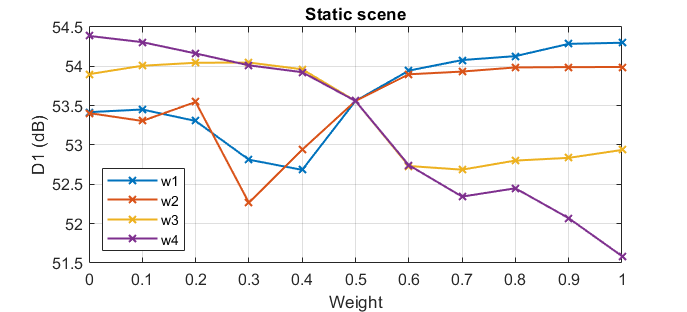} 
	\includegraphics[width=0.95\linewidth]{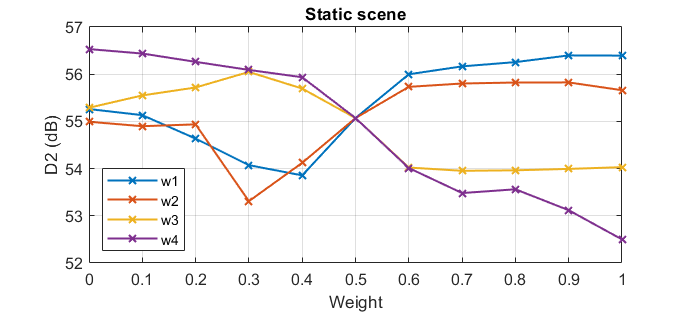} 
	\caption{Weight contribution on the final saliency map}
	\label{W}
\end{figure}
\begin{table*}[!htbp]
	\centering
	\begin{tabular}{|c | c | c |} 
		\hline 
		 \textbf{\footnotesize Variable}	&\textbf{\footnotesize Default Value} &  \textbf{\footnotesize Short description} \\  
		\hline\hline
		\thead{$k_n$} & \thead{6} & \thead{Number of neighboring vertices for the estimation of adjacency matrix $\mathbf{C}$ in Eq. \eqref{adjacency}, and point normals}\\ 
		\hline
		\thead{$e$} &  \thead{ $\begin{bmatrix}0 & 0 & 2 max{z} \end{bmatrix}^T$ } & \thead{User (or camera) position in 3D space }\\
		\hline
		\thead{$r$} &  \thead{ $\begin{bmatrix}0 & 0 & - 2 max{z} \end{bmatrix}^T$ } & \thead{Viewing direction }\\
		\hline
		\thead{$k_a$} & \thead{125} & \thead{Number of neighboring vertices in screen-space for the estimation of $a$ in Eq. \eqref{a_operator}} \\ \hline
		\thead{$k_g$} & \thead{25} & \thead{Number of neighboring points lying in an area and used for the estimation of matrices $\mathbf{E}$ and $\mathbf{G}$ in Eq. \eqref{NN1}, \eqref{NN2}} \\ 
		\hline 
		\thead{$s_0$} & \thead{$2 * mean( s_{11i})$} & \thead{Saliency threshold in Eq. \eqref{EQ}}\\
		\hline
		\thead{$k_c$} & \thead{0.01*n} & \thead{Number of the anchor points that corresponds to the $1\%$ of the total number of vertices $n$} \\ 
		\hline
		\thead{ $\begin{bmatrix} w_1 & w_2 & w_3 & w_4 \end{bmatrix}^T$} & \thead{ $\begin{bmatrix} 1 & 1 & 0.1 & 0.1 \end{bmatrix}^T$} & \thead{Weights for the combination of the saliency maps in Eq. \eqref{total_saliency}}\\
		\hline
	\end{tabular}
	\caption{Default values for parameters.}
	\label{table:values_of_parameter}
\end{table*}
The saliency threshold in Eq. \eqref{EQ} was set to $s_0 = 2mean( s_{11i})$ since curvature estimation is restricted to the most significant features of the first step. Anchor points are necessary because they minimize quantization errors during the reconstruction of the 3D PC, but their number should be insignificant in comparison to the size of the PC in order not to affect the compression ratio. For that purpose, we set $k_c = 0.01 n$ which means that anchor points correspond to the $1\%$ of the total number of vertices.
We also investigated the individual impact of each map on the final saliency,  in order to come up with the best strategy for defining the aforementioned weights of Eq. \eqref{total_saliency}. For that reason, each weight $w_i \ \forall \ i = 1,\cdots,4$ was set in the range $[0 - 1]$ while the remaining weights were given a fixed value i.e. $0.5$. We apply our compression scheme to the ``static scene" for each group of weights and asses the results with D1 and D2 metrics. The $s_{thresh}$ was changed in order to achieve the same bits per point (bpp) in each run, while the remaining parameters of our pipeline had their default values. In Fig. \ref{W}, we present the contribution of each weight to the reconstruction quality when the ``static scene" is compressed with $0.5$ bpp. As we expected, visibility and geometry saliency increase the overall reconstruction accuracy when multiplied with weights that are equal to 1. On the other hand, due to their subjective nature, focus and depth saliency contribute negatively to the extended saliency metric as the quality is reduced in big parts of the scene. Their presence on the final map is perceptually important since the quality is better distributed on the scene for small bit rates. For that reason, we assign small values to their weights that increase both metrics like $w_3 = w_4= 0.1 $. Table \ref{table:values_of_parameter} summarizes the default values that we used and a short description.
\begin{figure}[!htbp]
	\centering
	\includegraphics[width=0.9\linewidth]{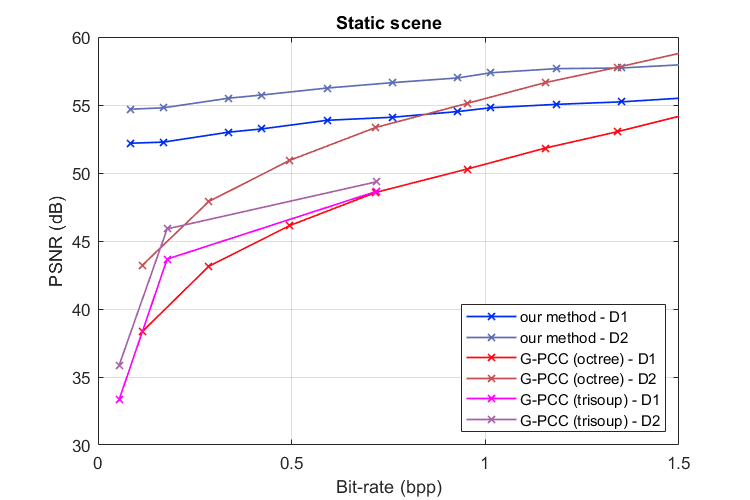} 
	\caption{Geometric point-to-point (D1) and point-to-plane (D2) PSNR comparisons. For our method we set $s_{thresh}$ in $[0.001,0.3]$, for G-PCC octree $\textit{positionQuantizationScale}$ in $[0.01,0.8]$, for G-PCC trisoup $\textit{trisoup\_node\_size\_log2} = 4,3,2$.}
	\label{fig:staticScene_D1_D2}
\end{figure}

\subsection{Quality Evaluation and Experimental Results}
\subsubsection{Objective Comparison}

\begin{figure*}[!h]
	\centering
	\begin{minipage}{0.33\textwidth}
		\centering
		\includegraphics[width=0.99\linewidth]{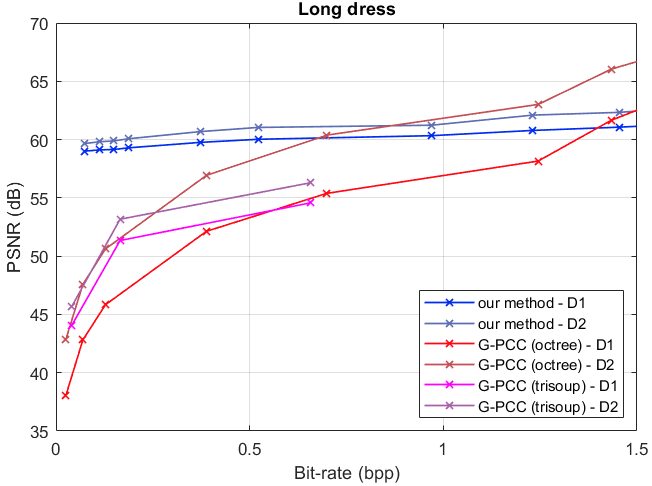} 
	\end{minipage}\hfill
	\begin{minipage}{0.33\textwidth}
		\centering
		\includegraphics[width=0.99\linewidth]{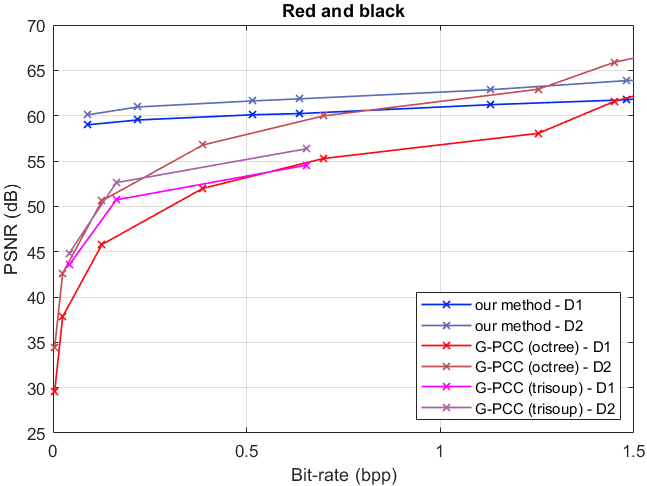} 
	\end{minipage}
	\begin{minipage}{0.33\textwidth}
		\centering
		\includegraphics[width=0.99\linewidth]{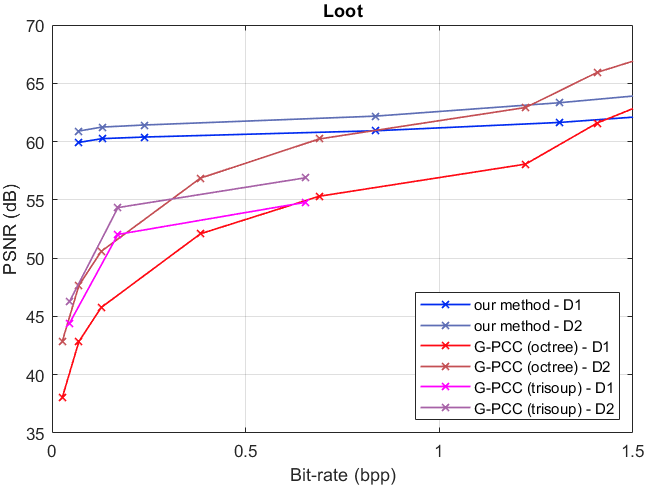} 
	\end{minipage}\hfill
	\begin{minipage}{0.33\textwidth}
		\centering
		\includegraphics[width=0.99\linewidth]{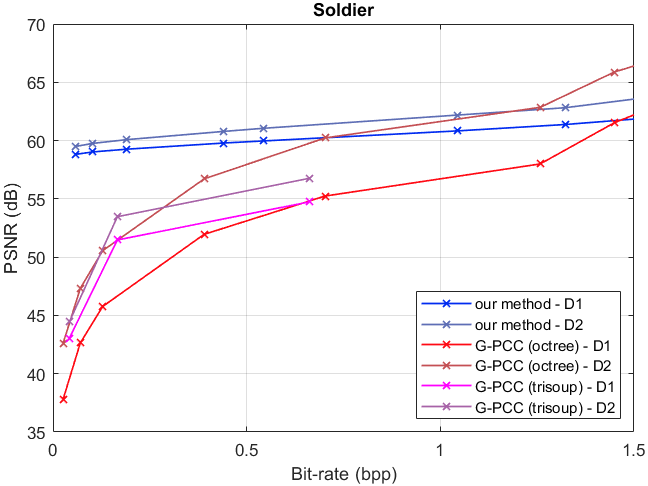} 
	\end{minipage}\hfil
	\begin{minipage}{0.33\textwidth}
		\centering
		\includegraphics[width=0.99\linewidth]{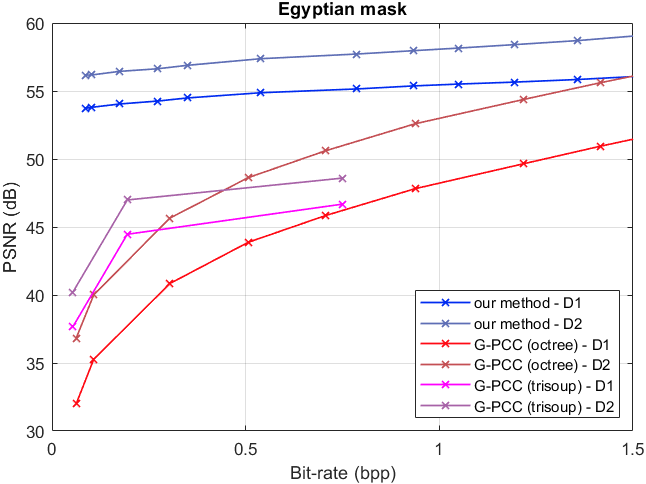} 
	\end{minipage}\hfill
	\begin{minipage}{0.33\textwidth}
		\centering
		\includegraphics[width=0.99\linewidth]{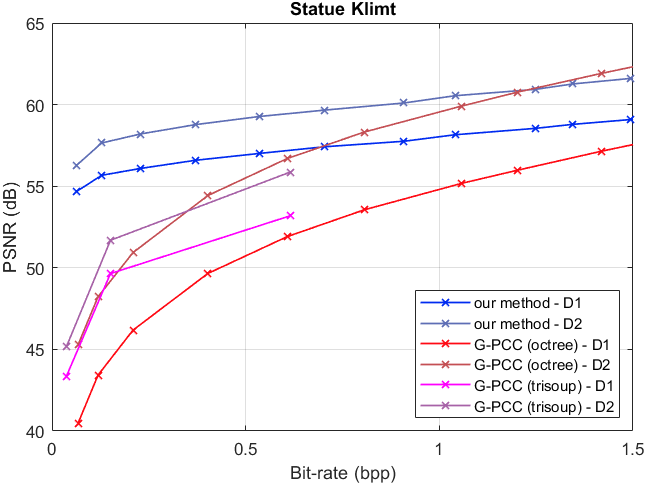} 
	\end{minipage}
	\begin{minipage}{0.33\textwidth}
		\centering
		\includegraphics[width=0.99\linewidth]{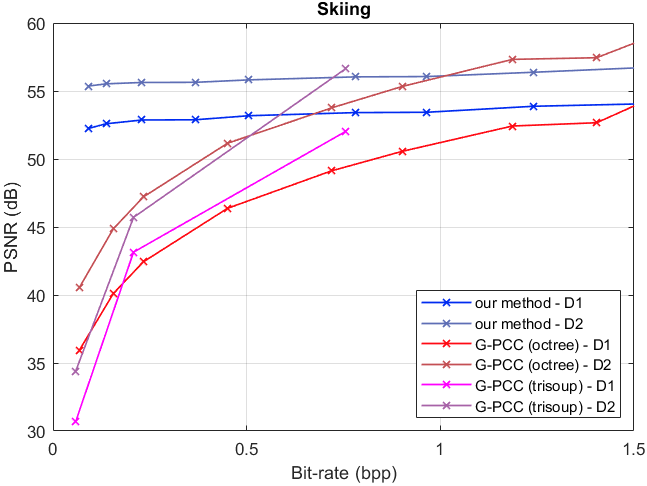} 
	\end{minipage}\hfill
	\begin{minipage}{0.33\textwidth}
		\centering
		\includegraphics[width=0.99\linewidth]{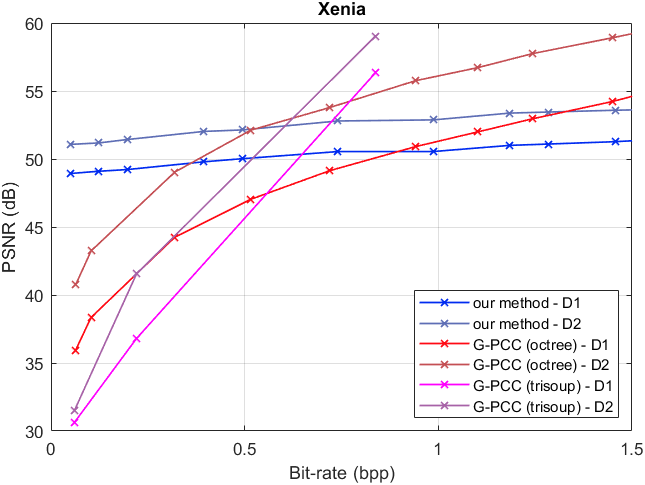} 
	\end{minipage}\hfil
	\caption{Geometric point-to-point (D1) and point-to-plane (D2) PSNR comparisons. For our method we set $s_{thresh}$ in $[0.001,0.3]$, for G-PCC octree $\textit{positionQuantizationScale}$ in $[0.01,0.8]$, for G-PCC trisoup $\textit{trisoup\_node\_size\_log2} = 4,3,2$.}
	\label{fig:D1_D2}
\end{figure*}

We compared our proposed PC codec with G-PCC using the latest test model category 13 (TMC13 v12.0 by MPEG), suitable for both static (category 1) and dynamically acquired (category 3) PCs. The TMC13 codec contains 2 geometry encoders (Octree and Trisoup) and 2 color encoders (Predlift and RAHT). Since our method encodes only geometry information we used the first two encoders. Although both compression frameworks reconstruct the whole PC scenes, for the comparisons we focused only on the visible points.
In order to compare our approach with G-PCC octree and G-PCC trisoup, we adjusted accordingly each codec's parameters for compressing the aforementioned PCs using the same bit rate range. For our method, we set $s_{thresh}$ of Eq. \eqref{quant-delta} from 0.3 to 0.001, leaving the parameters of Table \ref{table:values_of_parameter} as default. The values we adopted for G-PCC's coding parameters are those defined by MPEG in the so-called Common Test Conditions (CTC) document \cite{perry2020jpeg}. For some parameters we used different values for enforcing the desired bit rates. To be more specific,
for both G-PCC geometry encoders, the \textit{positionQuantizationScale} parameter was configured to determine the maximum voxel depth of the compressed PC. The \textit{trisoup\_node\_size\_log2} was additionally modified to define the size of the block to which the triangular soup approximation is applied. So, for G-PCC octree we set \textit{positionQuantizationScale} from 0.8 to 0.01 and for G-PCC trisoup we set \textit{trisoup\_node\_size\_log2} to 2, 3, 4, and \textit{positionQuantizationScale} to 1 for denser PCs such as the VCL/ITI’s dataset and to small values for sparser models such as 0.1 for static scene 1 and 0.08 for Statue Klimt. This downscaling was also applied in \cite{wang2019learned} for typically sparser but with higher precision models. Due to the fact that each model has different properties and structure, they require different ranges of the aforementioned parameters ($s_{thresh}$,  \textit{positionQuantizationScale},
\textit{trisoup\_node\_size\_log2}) in order to achieve a range of $0.02 - 1.6$ bits per point (bpp). The adopted objective quality criteria for the rate-distortion performance assessment are those used by MPEG, the aforementioned geometric point-to-point (D1) and point-to-plane (D2) PSNR metrics. Rate-distortion (RD) curves are presented in the Figs. \ref{fig:staticScene_D1_D2} and \ref{fig:D1_D2}. Table \ref{table:BD_psnr} presents the BD-PSNR against octree G-PCC for bit-rates below 1. Both the figures and the table indicate that the proposed method achieves significant quality results for minor bit rates. For bit rates less than one, the performance of our algorithm on average is better for $10.1756$ dB and $7.1748$ dB for D1 and D2 metrics respectively, compared to octree G-PCC. Notably for even smaller bit rates the results are more satisfactory. Since the differences in terms of quality change minimally for bigger bit rates, our method is ideal for aggressive compression ratios. 
\begin{table}[!htbp]
	\centering
	\begin{tabular}{|c|c|c|}
		\hline
		\multicolumn{1}{|c|}{\multirow{2}{*}{\small \textbf{\small Point cloud}}} & \multicolumn{2}{c|}{\small \textbf\small {BD-PSNR (dB)}}                          \\ \cline{2-3} 
		\multicolumn{1}{|c|}{}                                      & \multicolumn{1}{c|}{\small \textbf{\small D1}} & \multicolumn{1}{c|}{\small \textbf{\small D2}} \\ \hline \hline
		\small Long dress   & \small -10.3379   & \small -6.3756  \\ \hline
		\small Red and black  & \small -10.3415  & \small -6.9602                  \\ \hline
		\small Loot & \small -11.4340  & \small -7.6811                  \\ \hline
		\small Soldier  & \small -11.6418   & \small -7.7196                    \\ \hline
		\small Egyptian mask   & \small -13.7355  & \small -11.4026                    \\ \hline
		\small Statue Klimt   &  \small -9.1403 & \small -7.1775                      \\ \hline
		\small Skiing         & \small -8.2717  &  \small -6.2799                     \\ \hline
		\small Xenia         & \small -6.5454  & \small -3.8541                     \\ \hline
		\small Static scene & \small -9.0917  & \small -6.7946                     \\ \hline \hline
		\textbf{\small Average} & \small -10.1756 & \small -7.1748                \\ \hline
	\end{tabular}
	\caption{BD-PSNR D1 and D2 comparisons against G-PCC (octree) for bit-rates less than 1.}
	\label{table:BD_psnr}
\end{table}
\subsubsection{Qualitative Evaluation }
\begin{figure}[h]
	\centering
	\includegraphics[width=0.95\linewidth]{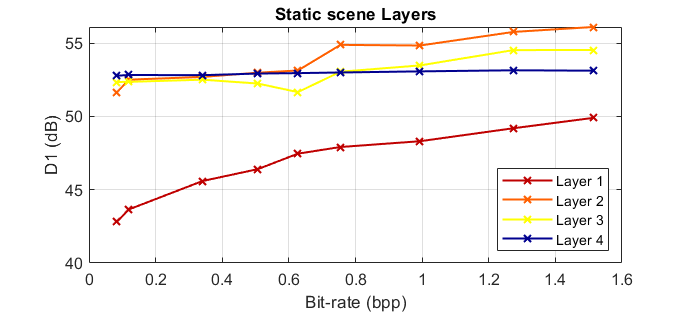} 
	\includegraphics[width=0.95\linewidth]{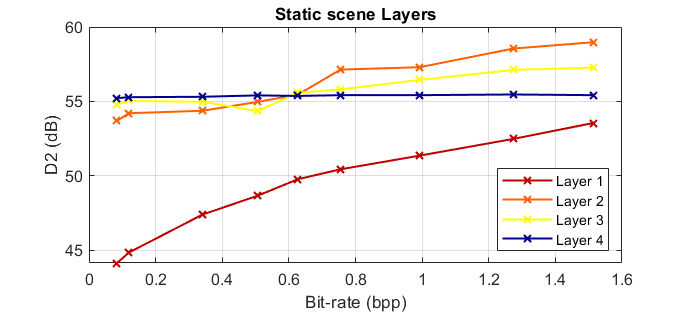} 
	\caption{PSNR comparisons for different layers.}
	\label{L}
\end{figure}

\begin{figure*}[!h]
	\includegraphics[width=0.99\linewidth]{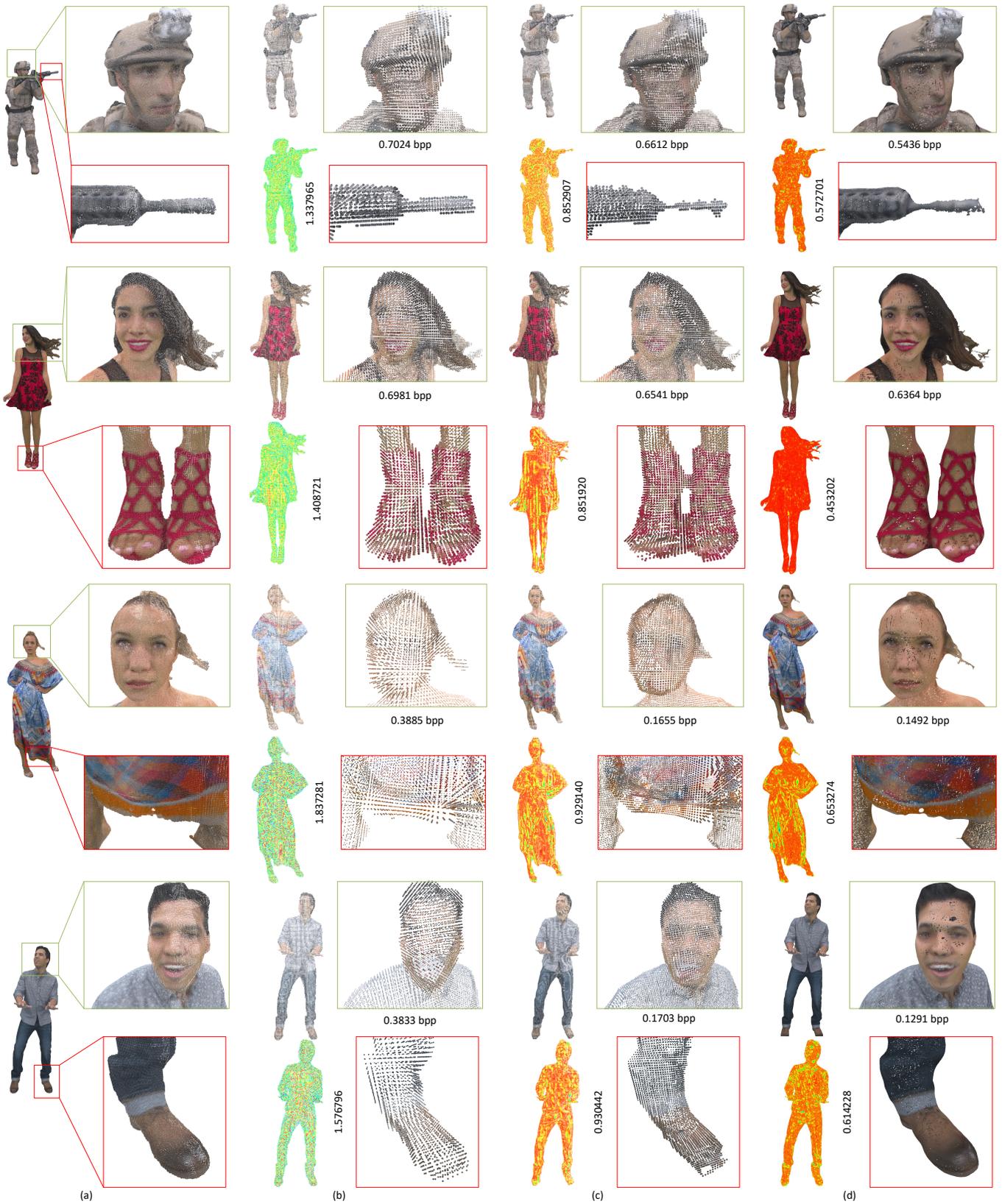}
	\caption{Indicative illustrations of (a) original models, and reconstructed results, heatmap visualizations of the Euclidean distance between the original and reconstructed models along with the mean Euclidean distance, using: (b) the  G-PCC octree approach, (c) the  G-PCC trisoup approach, (d) our approach.  }
	\label{reconstructed_results}
\end{figure*}

\begin{figure}[H]
	\begin{center}
		\includegraphics[width=0.95\linewidth]{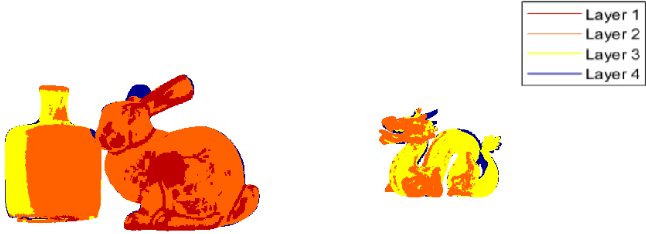}
	\end{center}
	\caption{Saliency-based layer separation.}
	\label{layers}
\end{figure}

In Fig. \ref{reconstructed_results}, we present the reconstructed results of our approach in comparison with the results of the other methods for the PC models ``Soldier", ``Red and black", ``Long dress" and ``Loot". Since our proposed scheme compresses only geometry data, we have given the original colors to the reconstructed models for visualization purposes. We also provide enlarged details of each one of the reconstructed models, for easier comparison among the methods and the corresponding bpp ratio. Additionally, we illustrate the heatmap visualization of each reconstructed model, highlighting in different colors (i.e., red color represents small differences while blue represents big differences) the Euclidean distance of each vertex with the corresponding vertex of the original model, and finally, we also provide the mean Euclidean distance. It is evident that our method preserves details in geometry for very small bit rates and provides good visual accuracy compared to G-PCC's octree and trisoup where the quality of reconstructed models is relatively poor. When trying to achieve such small bit rates G-PCC leads to high quantization errors. 
We further extend our studies by examining the performance of our proposed compression scheme in areas of distinct perceptual significance. To that end, we separated the input scene into 4 layers based on the extended saliency metric. The points in layer 1 have saliency $s > 0.7$, in 2 $ 0.4< s < 0.7$, in 3 $s < 0.4$ and the points of layer 4 are fully occluded ($s = 0$). Therefore, each layer is encoded with a different number of bits. 
In Fig. \ref{layers} we present the saliency-based layer separation for the static scene, and in Fig. \ref{L} the rate distortion curves for each layer. From these figures we note that layer 4 is almost unaffected by decreasing the bit rate, since it consists of points encoded with zero bits. In contrast to the other layers, layer 1 exhibits higher levels of distortion. This occurs because the areas with high geometric saliency represent high frequency information. 
In Table \ref{table:l}, we compare our saliency-aware encoding method with a scheme that assigns the same number of bits to each layer. By assigning more bits to layer 1, our method exploits the aforementioned characteristic and achieves better reconstruction results. 

\begin{table}[!ht]
	\begin{tabular}{|c|c|c|c|c|}
		\hline
		\small \textbf{Quantization}  & \small \textbf{Layer 1}      & \small \textbf{Layer 2}      & \small \textbf{Layer 3}      & \small \textbf{Layer 4}      \\ \hline \hline
		\small Saliency  & \small 49.1895 & \small 55.7560 & \small 54.5169 & \small 53.1400 \\ \hline
		\small Uniform  & \small 46.1024 & \small 53.2691 & \small 53.2331 & \small \small 53.1046 \\ \hline
	\end{tabular}
	\caption{D1 comparisons at 1.2 bpp.}
	\label{table:l}
\end{table}


\section{Conclusion}
In this work we presented a novel, geometry-based, end-to-end compression scheme for static PCs. Our proposed method highlights the most visually significant parts of the PC and compresses the position of each point based on an extended saliency metric that combines viewer’s relative position and geometric saliency. The quality reduction in perceptually insignificant parts of the scene adds a realistic sensation and is not noticeable by the user, even for aggressive compression rates.
Extensive assessment tests, performed with a dataset of PCs with different characteristics, verify the superiority of our approach for aggressive bit rates, as compared to two benchmarks of MPEG, namely G-PCC octree and G-PCC trisoup. Rate distortion curves and BD-PSNR table prove that our method is better for bit rates less than one. Qualitative tests have shown that for our method, the quality of the reconstructed PCs remains almost unaffected even for very small bit rates. The majority of geometric detail is lost for the baseline benchmark methods for small rates. So, we could extend our studies by integrating the saliency-aware encoding scheme to user-interactive rendering applications in order to increase compression efficiency in such scenes.
There is one major limitation of the proposed framework that could be addressed in future research. The reconstruction of the PC, based on solving a sparse linear system on the decoder side, leads to high execution times compared to SoA compression codecs. It is essential to focus future efforts on reducing execution time at the decoder using efficient schemes and parallelizable implementations in order to allow real-time performance on commodity hardware.

\footnotesize
\bibliographystyle{IEEEtran}
\bibliography{IEEEexample,evaluation,visibility,datasets} 



\end{document}